\documentclass[twocolumn]{autart}    
\usepackage{natbib}
\usepackage{har2nat}
\usepackage{graphicx}
\usepackage{amsmath}
\usepackage{amssymb}
\usepackage{bbm}
\usepackage{color}

\def\twon #1{\|#1\|}

\def\ra{\rightarrow}

\def\bN{\mathbb{N}}

\def\bR{\mathbb{R}}

\def\cE{\mathcal{E}}

\def\cG{\mathcal{G}}

\def\cP{\mathcal{P}}

\def\cV{\mathcal{V}}

\def\cX{\mathcal{X}}

\def\diag{\text{diag}}

\def \qed {\hfill \vrule height6pt width 6pt depth 0pt}
\def\bee{\begin{equation}}
\def\ene{\end{equation}}
\def\beq{\begin{eqnarray}}
\def\enq{\end{eqnarray}}

\begin{document}
\begin{frontmatter} 

\title{Cooperative Source Seeking via Networked Multi-vehicle Systems\thanksref{footnoteinfo}} \thanks[footnoteinfo] {This research was supported in part by the National Key Research and Development Program of China under Grant No.2016YFC0300801, and National Natural Science Foundation of China under Grants No.41576101 and No.41427806. } \author{Zhuo~Li},
\ead{lizhuo16@mails.tsinghua.edu.cn}
 \author{Keyou~You\corauthref{cor}},
 \ead{youky@tsinghua.edu.cn}
\author{Shiji~Song}
\ead{shijis@tsinghua.edu.cn}
\corauth[cor]{Corresponding author}\address{Department of Automation, and BNRist, Tsinghua University, Beijing, 100084, China.}

\begin{abstract} 
This paper studies the cooperative source seeking problem via a networked multi-vehicle system. In contrast to existing literature, {the multi-vehicle system} is controlled to the {source} position that maximizes aggregated {\em multiple}  unknown scalar fields and each sensor-enabled vehicle only {samples} measurements of {\em one} scalar field. Thus, a single vehicle is unable to localize the source and has to cooperate with its neighboring vehicles. By jointly exploiting the ideas of the consensus algorithm and the stochastic extremum seeking (ES), this paper proposes novel distributed stochastic ES controllers, which are gradient-free and do not need any {absolute information},  such that the multi-vehicle system simultaneously approaches the source position.  The effectiveness of the proposed controllers is proved for quadratic scalar fields. Finally, illustrative examples are included to validate the theoretical results.   

\end{abstract}
\end{frontmatter}
\section{Introduction}

This paper is concerned with the design of distributed controllers to drive a multi-vehicle system to approach a source position of interest, which has great significance in various applications, such as environmental monitoring \citep{dhariwal2004bacterium}, odor source detection \citep{gao2016detection}, acoustic source localization \citep{zhao2016acoustic} and pollution sensing \citep{gao20163d}. Consider the problem of seeking an indoor fire source, where there are multiple indoor positions having either the highest temperature or the highest toxic gas concentration but only the position of the fire source attains the highest values of the both fields. {Consequently}, it is unable to localize the fire source by sensing only one of the two scalar fields. There are also examples that the position of interest may not be a maximum of any sensed scalar field. {Based on these observations}, we are interested in the complex environment where the source position maximizes the \emph{aggregated multiple unknown scalar fields}, which is different from existing works {exploring only one scalar field} \citep{zhang2016extremum, durr2017extremum, lin2017stochastic}.

Our first challenge lies in the unknown distribution of any scalar field, i.e., any sensor cannot measure a continuum of the scalar field. Hence the distributed optimization algorithms explicitly using gradients  cannot be directly applied here \citep{wang2010control, gharesifard2014distributed, you2019distributed}. To solve it, we adopt the stochastic extremum seeking (ES) \citep{manzie2009extremum} to estimate local gradients by using samples of the sensed scalar fields, which is completed in the associated vehicle by superimposing {a stochastic excitation signal}. Moreover, the ES method does not require absolute position information.

The second is that each vehicle of this work is only able to sense one scalar field. In this case, {seeking the source position for multiple scalar fields} needs multiple vehicles and their cooperation. Although the networked multi-vehicle system has been employed in \citet{frihauf2014single, khong2014multi, bri2016distributed, turgeman2018multiple}, the seeking position therein is the source of only one scalar field and all the vehicles take samples from the same field. Thus, the cooperation in their works is not indispensable.

By jointly using stochastic ES and the cooperation among vehicles, this work proposes distributed stochastic ES (DSES) controllers to drive all vehicles to  simultaneously approach the source position. In the literature, cooperative ES has been used for social games in \citet{menon2014collaborative, dougherty2017extremum, vandermeulen2018discrete, guay2018distributed} and the resource allocation problem in \citet{poveda2013distributed}.  In their works, the objective of each agent is to reach the social equilibrium or compute its optimal resource via cooperation. Clearly, they cannot apply to our problem, since we require each vehicle to reach consensus at the same source position.

Our problem setting is closely related to \citet{ye2016distributed,kvaternik2012analytic, michalowsky2017distributed}. In \citet{ye2016distributed}, a consensus-based ES algorithm is developed to solve a saddle point problem. Since their algorithm does not involve the agent dynamics, it is unclear how to extend it to dynamical vehicle models, e.g., the unicycle model in \citet{li2015distributed}. Moreover, the excitation of ES therein is based on deterministic signals, which should be orthogonal among agents. {This renders it not as simple as using the stochastic signals for implementation in the multi-vehicle network.} In \citet{kvaternik2012analytic, michalowsky2017distributed}, authors show the existence of a distributed ES controller to find the position of interest in the deterministic regime, {but do not provide} an explicit ES controller.

We prove the effectiveness of DSES controllers for the quadratic scalar fields by the stochastic averaging theory,  {and conclude that the DSES controllers might also work for the non-quadratic case by simulations}.  A conference version of this work has been presented in \citet{li2018distributed} where the DSES controller is given for a special case that the position of interest simultaneously maximizes all the local {scalar fields}.

The rest of this paper is organized as follows. In Section \ref{sec:formulation}, we describe the cooperative seeking problem by using a group of networked vehicles. In Section \ref{sec:integrator}, we propose {DSES} controllers in both undirected and directed interaction graphs and prove their effectiveness by the stochastic averaging theory. Illustrative examples are provided in Section \ref{sec:exam} and some remarks are drawn in Section \ref{sec:conclusion}.

{\bf Notation}: Throughout this paper, any notation with a subscript $i$ represents that of vehicle $i$, e.g., $x_i \in \bR^m$, and $x=[x_1', x_2', \ldots, x_n']'\in \bR^{m n}$ for the networked multi-vehicle system. $O(\alpha)$ denotes the infinitesimal of the same order as a scalar $\alpha$, i.e., $\lim_{\alpha\rightarrow 0}O(\alpha)/\alpha=c<\infty$ with $c \not = 0$. $\|\cdot \|$ denote the Euclidean norm for a vector or a matrix and $\otimes $ denote the Kronecker product. 
 
\section{Problem Formulation}
\label{sec:formulation}

In this section, we explicitly describe our cooperative source seeking problem by using the networked multi-vehicle system, where each vehicle is embedded with only one sensor to measure the strength of one scalar field and has to cooperate with its neighboring vehicles.  

\subsection{The cooperative source seeking problem}   \label{sub:exam}
There are $n$ networked autonomous vehicles, each of which has only one sensor to measure the signal strength {a scalar field} $f_i(z): \bR^m \to \bR, i=1,2, \cdots n$ at the position $z \in \bR^m$. The task of the multi-vehicle system is to autonomously approach the source position $z^*$ that maximizes the sum of $f_i(z)$, i.e., 
\bee \label{eqn:opt1}
z^*\in \arg\max_z F(z):=\sum\nolimits_{i=1}^n f_i(z),
\ene
where $\arg \max_{z} F(z)$ denotes the set of optimal points that maximize the aggregated multiple unknown scalar fields. 

Since the value of $f_i(\cdot)$ is the {\em only} accessible information from the sensed scalar field for the $i$-th vehicle, each vehicle has to cooperate with others to complete the seeking task in (\ref{eqn:opt1}). This problem setup is essentially motivated by two notable examples.

\begin{exmp} \label{exmp:1}
Consider an indoor fire source seeking problem. The fire source $z^*$ is the position of our interest, and is the unique point that simultaneously attains the highest temperature and toxic gas concentration, i.e., 
\bee \label{eqn:p0}
z^*\in \bigcap\nolimits_{i=1}^2 \arg \max \limits_{z} f_i(z),
\ene 
where $f_1(z)$, $f_2(z)$ denote the temperature and the toxic gas concentration at the position $z \in \bR^m$, respectively.

To approach the fire source $z^*$, there are two autonomous vehicles embedded with a temperature sensor and a gas sensor, respectively. Due to the complex sensing environment, it is possible that each $f_i(z)$ contains multiple or an infinite number of maximum points. Then, any vehicle cannot guarantee to exactly find the fire source $z^*$ and has to cooperate with others. One can easily show that the multi-objective problem \eqref{eqn:p0} {is a special case of} the cooperative seeking problem \eqref{eqn:opt1}.  \qed
\end{exmp}

\begin{exmp} \label{exmp:2}
Consider the following dynamical process 
$$z(k+1)=Az(k)$$
where $A\in \bR^{m\times m}$ and $z(k)\in\bR^m$ denote the transition matrix and the state at time step $k$, respectively. Our objective is to recover the initial state $z(t_0)$ by using measurements from multiple sensors.

At each time step $k$, the $i$-th vehicle is able to take the measurement 
$$
{m_i(k)}=C_iz(k),
$$
where the system $(C_i,A)$ is {\em not observable} for any $i$,  i.e., the observability Gramian $\Phi_i:=\sum_{k=0}^{m-1}(A')^kC_i'C_i A^k$ is rank deficient {\citep{chen1998linear}}. {Therefore}, it is impossible to recover the initial state $z(t_0)$  by only using measurements of any single vehicle. Now, suppose that the multi-vehicle system is jointly observable, i.e. {the system} $(C, A)$ is observable with $C=(C_1', \ldots, C_n')' $. Then, the multi-vehicle system is able to cooperatively complete the recovering task.  
 
To elaborate it, define the local objective function as
\bee\label{localmea}
f_i(z)=-\sum\nolimits_{k=0}^{m-1}\twon{m_i(k)-C_i A^kz}^2.
\ene
It is easy to show that $\arg\max_z f_i(z)=z(t_0)+\text{Null}(\Phi_i)$ where $\text{Null}(\Phi_i)$ denotes the null space of $\Phi_i$. 
Since $(C_i,A)$ is not observable, then $\text{Null}(\Phi_i)$ is a non-trivial subspace of $\bR^m$. Therefore, $z(t_0)$ is not the unique element of $\arg\max_z f_i(z)$. That is, $z(t_0)$ is unable to be recovered by only using the $i$-th vehicle's measurements. {Similarly}, we can show that $\arg\max_z F(z)=z(t_0)+\text{Null}(\sum_{i=1}^n\Phi_i)$. 
Since $(C, A)$ is observable, then $\sum_{i=1}^n\Phi_i$ is non-singular, which in turn implies that $\arg\max_z F(z)=z(t_0)$. That is,  $z(t_0)$ is the unique element of  $\arg\max_z F(z)$, and {is able to be recovered by solving the cooperative seeking problem (\ref{eqn:opt1}) with $f_i(z)$ given in (\ref{localmea}).} \qed
\end{exmp}

In the above examples, each set of {\em local} optimal points $\arg\max_z f_i(z)$ may contain multiple elements, and we are only interested in the one lying in their intersection, which clearly maximizes $F(z)$ the sum of all the local objective functions. It should be noted that the cooperative seeking problem \eqref{eqn:opt1} also includes the case where {the maximum point of $F(z)$, i.e., the source position $z^*$,} may not maximize any $f_i(z)$. In both cases, a local objective function $f_i(z)$ can only offer limited information on the source position $z^*$, the localization of which obviously requires the cooperation among vehicles. 

\subsection{Networked multi-vehicle systems}

The interactions (cooperations) between vehicles are modeled by a graph $\mathcal{G}=\{\mathcal{\cV,\cE}\}$, where $\mathcal{V}=\{1,2,\ldots,n\}$ is the index set of nodes (vehicles) and $\cE\subseteq \cV\times\cV$ is the set  of the interaction edges between vehicles. Node $i$ can measure its relative position to that of node $j$ if and only if $(i,j)\in\cE$. The set of neighbors of node $i$ is denoted by $\mathcal{N}_i=\{j\in \mathcal{V}:(i,j)\in \mathcal {E}\}$. A path from node $j_0$ to node $j_k$ is a set of distinct nodes $\{ j_0,j_1, \ldots, j_k\}$ such that $(j_{i-1},j_{i})\in \mathcal{E}$ for all $i\in\{1, 2, \ldots,k\}$. If any two nodes can be connected via a path, then $\cG$ is strongly connected.  Let the adjacency matrix $[a_{ij}]_{n\times n}$ be defined such that $a_{ij}>0$ if $(i,j)\in \mathcal{E}$ and $a_{ij}=0$ otherwise. The associated Laplacian matrix is $L=[l_{ij}]_{n\times n}$ where $l_{ii}=\sum_{j=1}^n a_{ij}$ and $l_{ij}= -a_{ij} $ for $i \not = j$, and $1_n$ is the unique solution (within a multiplier) of $Lx=0$ if $\cG$ is strongly connected \citep{ren2008distributed}. If $L=L^T$, then $\cG$ is an undirected graph.  {Clearly, the connectivity of $\cG$ is necessary for our problem. }
\begin{assum} \label{assum_graph} $\cG$ is strongly connected.
\end{assum}

The networked $n$ autonomous vehicles are modeled by single integrators
\bee \label{eqn:integrator}
\dot z_i= u_i, i \in \mathcal{V},
\ene
where $z_i(t)\in \mathbb{R}^m$ and $u_i(t) \in \mathbb{R}^m$ represent the position and the control input of the $i$-th vehicle at time $t$, respectively. When it is clear from the context, we drop the dependence of the time index $t$ for ease of notations.

\subsection{The objective of this work}
The objective of this work is to design  distributed controllers for the networked multi-vehicle system to {{\em simultaneously}} approach the source position $z^*$ of (\ref{eqn:opt1}) under the following constraints:

\begin{enumerate}
\renewcommand{\labelenumi}{\rm(\alph{enumi})}
\item \label{con:a} Each vehicle $i$ is only able to obtain the numerical value of $f_i(z_i)$ at its current position $z_i$.
\item Each vehicle $i$ can only measure its relative positions to its neighbors and {has no access to} its absolute position in the global position system (GPS).
\end{enumerate}

Under the first constraint, gradient-based methods cannot be directly applied to solve the cooperative source seeking problem (\ref{eqn:opt1}). We adopt the stochastic ES method \citep{liu2010stochastic} to design a gradient-free controller for each vehicle $i$, which is rigorously proved for the quadratic $f_i(\cdot)$. Although the ES method has been widely applied to solve source seeking problems, the number of scalar fields is mostly restricted to one, i.e. $n=1$ in \eqref{eqn:opt1}. In these cases, a single vehicle is sufficient to complete the seeking task, and there is no need of vehicles' cooperation. In contrast, the cooperation among vehicles is indispensable in this work, and is achieved by using consensus algorithms \citep{ren2008distributed}.

Under the second constraint, distributed controllers are designed by only using the relative positions to its neighbors, which is particularly useful in the GPS-denied environment, e.g., the indoor fire source seeking. This is essentially motivated by the observation that many sensors, e.g., the acoustic sensor and the vision sensor, can easily measure the relative positions between two vehicles, while it is difficult to obtain the vehicle's GPS information. From this point of view, our controller preserves the advantage of the ES method without using any absolute position information. 

It is {worth} mentioning that if the vehicle can only take a noisy measurement {$f_i(z)+\varepsilon_i(z)$ at the position $z$, where} $\varepsilon_i(z)$ is an additive white noise and is spatially independent, our major results still hold.  

\section{Distributed Stochastic ES Controller Design} 
\label{sec:integrator}

In this section, we design distributed stochastic ES (DSES) controllers for the networked vehicles and prove that the multi-vehicle system converges to the source position $z^*$ in (\ref{eqn:opt1}). We first consider undirected graphs and then extend to directed graphs. 

\subsection{The DSES controller for undirected graphs}
\label{sub:int_v}

\begin{figure}
  \centering
\includegraphics[width=7cm]{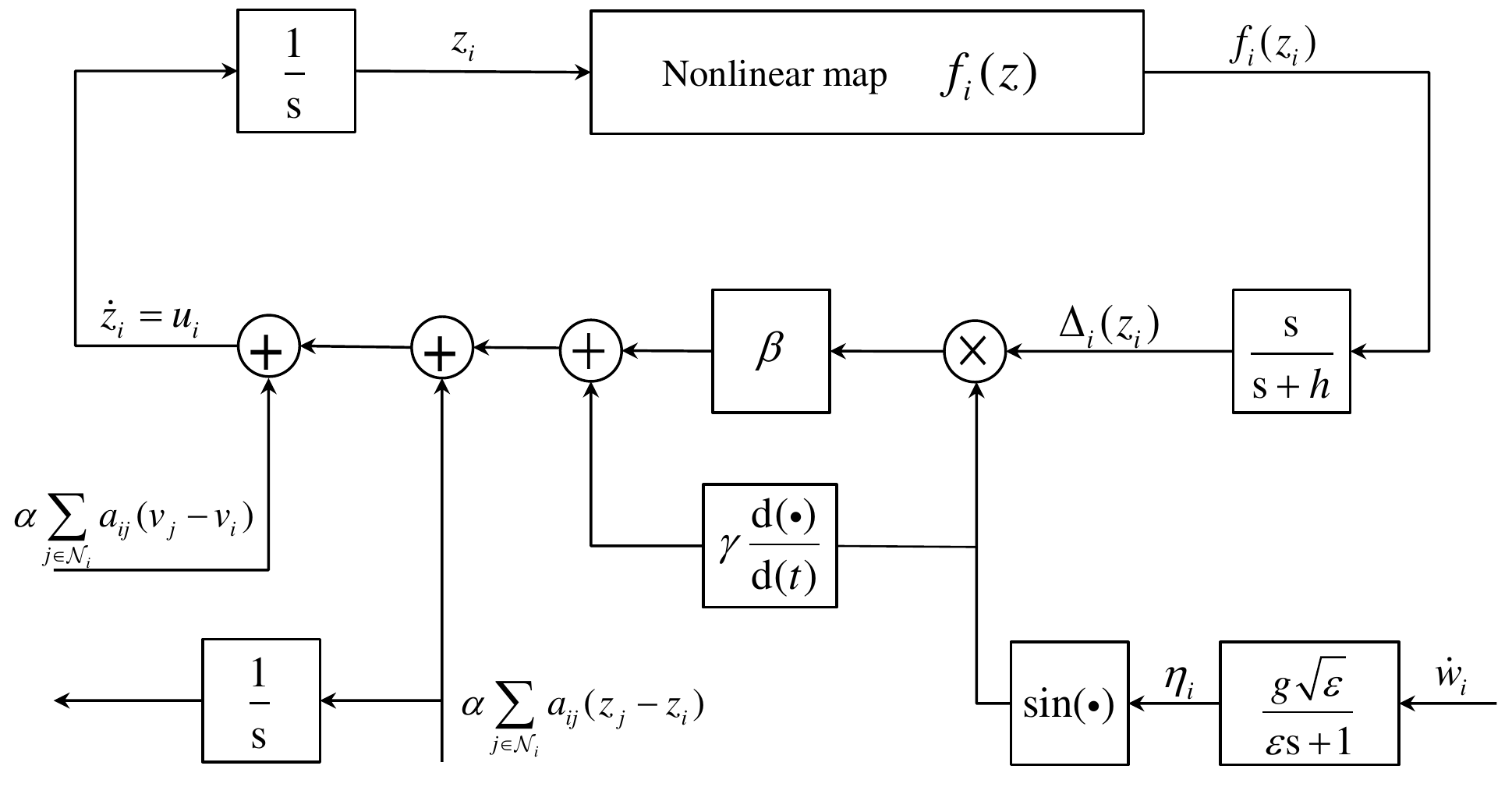} 
\caption{The DSES controller \eqref{eqn:control_1} for undirected graphs.}
\label{single_integrator_v}
\end{figure}

If $\cG$ is undirected, the DSES controller for the $i$-th autonomous vehicle is devised as
\bee \label{eqn:control_1}
\left \{
\begin{aligned}
u_i&=\alpha \sum\nolimits_{j \in \mathcal{N}_i}a_{ij}(z_j-z_i+v_j-v_i)\\
&~~~+\beta \sin(\eta_i) \Delta_i(z_i)+\gamma {\text{d} \text{sin}(\eta_i)}/{\text{d}t},\\
\dot v_i&= \sum\nolimits_{j \in \mathcal{N}_i}a_{ij}(z_i-z_j),
\end{aligned}
\right.
\ene
where $\alpha, \beta, \gamma \in \bR_{+}$ are positive control parameters, $v_i \in \bR^{m}$, $\Delta_i(z_i)$ is the output of a washout filter $\frac{\text{s}}{\text{s}+h}$ under the input signal $f_i(z_i)$, and $\sin(\eta_i)$ is the sinusoid of a stochastic excitation signal $\eta_i \in \bR^{m}$, which is the state output of the following diffusion process
\bee \label{eqn:eta}
\text{d}\eta_i=-\frac{1}{\epsilon}\eta_i \text{d}t + \frac{g}{\sqrt{\epsilon}}\text{d}w_i,
\ene
where $\epsilon\in(0,\epsilon_0]$ is a given parameter for some fixed $\epsilon_0>0$, $g$ is any positive parameter, and $w_i$ is a standard $m$-dimensional Brownian motion \citep{oksendal2003stochastic}. Note that $w_i$ is generated independently of $w_j$ for any $j\neq i$. The last term ${\text{d} \sin(\eta_i)}/{\text{d}t}$ is the Ito derivative of $\sin(\eta_i)$ to persistently excite the system. In \citet{manzie2009extremum, frihauf2014single}, the selection of $\gamma, h, \epsilon_0, g$ is discussed for the standard ES method and is not repeated here.  See Fig. \ref{single_integrator_v} for the {DSES} controller of each vehicle. 

The cooperation among vehicles is exploited in the first term of  $u_i$ in \eqref{eqn:control_1}, where $\sum_{j \in \mathcal{N}_i}a_{ij}(z_i-z_j)$ is also known as the consensus term \citep{ren2008distributed} and $v_i$ is its integration. This term only relies on the relative positions of the $i$-th vehicle to its neighbors and is to coordinate vehicles. Thus, each vehicle utilizes not only its own local measurements, but also its neighbors' trajectories, which is exactly the advantage of cooperation. This idea is significantly different from \citet{vandermeulen2018discrete, dougherty2017extremum, guay2018distributed} where a consensus algorithm is used to estimate the sum of local objective functions $f_i(\cdot)$.

To implicitly estimate the gradient of $f_i(\cdot)$, we adopt the stochastic ES technique in \citet{liu2010stochastic}. The last two terms of $u_i$ in \eqref{eqn:control_1} are the approximated gradient and the stochastic excitation signal, respectively. {Note that it is always difficult for the deterministic ES to satisfy orthogonality requirements for a large number of networked vehicles. Excitation signals} using the Brownian motion $w_i$ for the stochastic ES are much easier to implement as we only require their independence. 

Overall, the DSES controller \eqref{eqn:control_1} jointly exploits the ideas from the consensus algorithm and the stochastic ES technique. From this perspective, the strictly positive parameters $\alpha$ and $\beta$  balance the importance of the consensus and the stochastic ES terms. Specifically, if $\alpha$ is relatively large, the vehicles tend to reach consensus faster, otherwise they tend to be attracted to their own individual sets of optimal points, i.e., $\arg\max_z f_i(z)$. This has been validated in the simulations. Roughly speaking, $\alpha$ can be interpreted as the rate of learning neighboring vehicles' behaviors and $\beta$ is the rate of learning its local objective function. To localize the source position $z^*$, both rates are essential and cannot be neglected. 

\subsection{{Convergence} analysis} 
We {establish the convergence} of the networked $n$ autonomous vehicles \eqref{eqn:integrator} with the DSES controller \eqref{eqn:control_1} under a similar assumption as \citet{liu2010stochastic}. 

\begin{assum} \label{assum_obj} The objective function $f_i(z)$ in (\ref{eqn:opt1}) is quadratic\footnote{For the non-quadratic case, it serves as a local quadratic approximation and the {convergence} results hold in the local sense.}, i.e. 
\bee \label{eqn:opt_ig}
f_i(z)=-\frac{1}{2} z' H_i z+b_i' z + c_i, \forall i\in\cV,
\ene 
where $H_i\in\bR^{m\times m}$ is positive semi-definite. Moreover, $\sum_{i=1}^nH_i$ is strictly positive definite. 
\end{assum} 

Though our theoretical result is established for the quadratic case, simulations in Section \ref{sec:exam} indicate the applicability of the proposed DSES controller to non-quadratic cases.

{Let $e_i=f_i(z_i)-\Delta_i(z_i)-f_i(z^*)$.} Inserting the DSES controller \eqref{eqn:control_1} to the vehicle's dynamical equation \eqref{eqn:integrator} leads to the closed-loop system
\bee \label{eqn:closed_loop}
\left \{
\begin{aligned}
\text{d}z_i&=\alpha \sum\nolimits_{j \in \mathcal{N}_i}a_{ij}(z_j-z_i+v_j-v_i)\text{d}t \\
&~~~+\beta \sin(\eta_i) \Delta_i\text{d}t +\gamma {\text{d} \text{sin}(\eta_i)},\\
\text{d} v_i&= \sum\nolimits_{j \in \mathcal{N}_i}a_{ij}(z_i-z_j) \text{d}t,\\
\text{d}e_i&=h\Delta_i\text{d}t,\\
\end{aligned}
\right.
\ene
where we adopt $\Delta_i$ to denote $\Delta_i(z_i)$ for notational simplicity.

\begin{prop} \label{prop:int_v}
Given an undirected graph $\cG$, consider the networked $n$ autonomous vehicles \eqref{eqn:integrator} under the DSES controller \eqref{eqn:control_1}. Let $\tilde{z}_i(t)=z_i(t) -\gamma \sin(\eta_i) -z^*$ where  $z^*$ is the source position defined in \eqref{eqn:opt1} and suppose that Assumptions \ref{assum_graph} and \ref{assum_obj} hold. Then, there exists a positive constant $\rho_1$ and a function $T_1(\epsilon): (0,\epsilon_0)\to \bN$ such that 
for any $\delta>0$ and bounded initial condition (i.e. $\twon{z_i(t_0)}<+\infty$, $\twon{v_i(t_0)}<+\infty$ for all $i\in\cV$), it holds that $\forall i \in \cV$,
\bee \label{eqn:con_cor}
\lim_{\epsilon \to 0} \cP\left\{\| \tilde z_i(t)\| \leq  \rho_1 \exp({-\lambda_1 t}) +\delta, \forall t \in [t_0,T_1(\epsilon)] \right\}=1,
\ene
where $\lim_{\epsilon\ra 0}T_1(\epsilon)=+\infty$ and $\lambda_1>0$ is the smallest eigenvalue of the following positive definite matrix
$$M:=\alpha (L \otimes I_m) + \kappa \beta H^d$$
{with $\kappa= \frac{\gamma}{2} (1-\exp(-g^2))$ and $H^d=\diag (H_1, \ldots, H_n)$.}
\end{prop}

By \eqref{eqn:con_cor}, the DSES controller is able to drive the multi-vehicle system to the neighborhood of the source position $z^*$ with an error size {$O(\gamma)+O(\delta)$} in probability. {Since $\delta$ is an arbitrarily small constant, it follows that the parameter $\gamma$ controls the distance of the vehicle's final position to the source position $z^*$.}

The exponential convergence in \citet{liu2010stochastic} for a single field still holds in the present problem. The major difference is that the rate here also depends on the interaction graph among vehicles. In view of the exponent $\lambda_1$, we conclude that the larger the {control parameters} $\alpha, \beta, \gamma$ and $g$, the faster the convergence rate of {the position error} $\tilde{z}_i(t)$ in the continuous-time regime. However, this does not apply to the discretized system in application. Particularly,  if { the control parameter} $\alpha, \beta, \gamma$ or $g$ is too large, it might lead to divergence.

The argument $\epsilon$ in $T_1(\epsilon)$ is exactly the parameter $\epsilon$ in the stochastic excitation signal $\eta_i$. That is, the convergence depends on the stochastic excitation signal. A sufficiently small $\epsilon$ is required to guarantee the reliability of the convergence.

{\em Proof of Proposition \ref{prop:int_v}}:
By the dynamical equation in \eqref{eqn:closed_loop}, we obtain the following dynamics\footnote{The derivative is interpreted as the Ito derivative, which is clear from the context.}
\beq\label{eqn:error}
\left\{
\begin{aligned}
\dot{\tilde z}_i&=\alpha \sum\nolimits_{j \in \mathcal{N}_i}a_{ij} \big( \tilde z_j+\gamma \text{sin}(\eta_j)-\tilde z_i -\gamma \text{sin}(\eta_i) \\
&~~~~~~~~~~~~~~~~~~+ v_j- v_i \big)+\beta \sin(\eta_i) \Delta_i ,\\
\dot{ v}_i&= \sum\nolimits_{j \in \mathcal{N}_i}a_{ij}(\tilde z_i+\gamma \text{sin}(\eta_i)-\tilde z_j -\gamma \text{sin}(\eta_j)), \\
\dot e_i&=h\Delta_i.  \\
\end{aligned}
\right.
\enq
The  proof is completed via  three steps.

{\em Step 1: Stochastic average system of \eqref{eqn:error}.}   

Let {the excitation signal} $\eta_i(t)=\chi_i(t/\epsilon)$ and substitute it into \eqref{eqn:error}. {The average of $\dot {\tilde z}_i$ is defined as
\beq \label{eqn:ave_1}
\begin{aligned}
\dot {\tilde z}_i^{a} =& \lim_{T \to +\infty} \frac{1}{T} \int_{t_0}^{T+t_0} \alpha \sum\nolimits_{j \in \mathcal{N}_i}a_{ij}\big(\tilde z_j+\gamma \text{sin}(\chi_j)\\
& - \tilde z_i-\gamma \text{sin}(\chi_i)+ v_j- v_i\big)+\beta \sin(\chi_i) \Delta_i dt,
\end{aligned}
\enq
where $t_0 \ge 0$ and $\chi_i$ denotes $\chi_i(t/\epsilon)$ for ease of notations. To compute the average in \eqref{eqn:ave_1}, we adopt $\varpi_i(t)={1}/{\sqrt\epsilon}\text{d}w_i(\epsilon t)$ to denote a standard Brownian motion and obtain that $ \text{d}\chi_i(t)=-\chi_i(t)\text{d} t+g\text{d} \varpi_i(t) $ from the diffusion process \eqref{eqn:eta}. Clearly, $\{\chi_i(t)\}_{t\ge t_0}$ is an ergodic Ornstein-Uhlenbeck process and has an invariant distribution $\mu (\text{d}s)={1}/({\sqrt \pi g})\exp(-{s^2}/{g^2})\text{d}s$. It follows from the ergodic theorem \citep{ash2000probability} that 
\bee \label{eqn:ergodic}
\lim_{T \to +\infty} \frac{1}{T} \int_{t_0}^{T+t_0} \sin(\chi_i(t)) \text{d}t=\int_\bR \sin(s)\mu (\text{d}s) 
\ene 
almost surely, which is uniform in $\forall t_0 \ge 0$.} Moreover, it holds that
\bee \label{eqn:tri1}
\int_{\bR}\sin^k(s)\mu (\text{d}s)=\left\{
\begin{array}{ll}
0,&k=1,3,\\
\frac{1}{2}(1-\exp(-g^2)),& k=2,
\end{array}
\right.\\
\ene
by \citet[Section III]{liu2010stochastic}, and
\beq \label{eqn:tri2}
\begin{aligned}
&\int_{\bR^2}\sin^{k} (s) \sin^{l} (t)\mu(\text{d}s) \mu(\text{d}t)=\frac{1}{\pi g^2}\\
&~~~\int_{\bR}\sin^{k}(s)\exp(-\frac{s^2}{g^2}) (\text{d}s) \int_{\bR}\sin^{l}(t)\exp(-\frac{t^2}{g^2}) (\text{d}t).
\end{aligned}
\enq
{Moreover, $\Delta_i$ in \eqref{eqn:ave_1} can be expressed by the definition of $e_i$,} i.e.,
\bee \nonumber
\begin{aligned}
\Delta_i &=-\frac{1}{2}\tilde{z}_i' H_i \tilde{z}_i -\frac{\gamma^2}{2}{\sin(\chi_i)}' H_i \sin(\chi_i) -\gamma\sin(\chi_i )' H_i \tilde{z}_i\\
&~~-\gamma \sin(\chi_i )' H_i z^* - \tilde{z}_i' H_i z^* +b_i' (\tilde{z}_i+ \gamma \sin(\chi_i)) -e_i.\\
\end{aligned}
\ene 
Thus, together with the ergodicity of $\chi_i$ in \eqref{eqn:ergodic} and the relations in \eqref{eqn:tri1}-\eqref{eqn:tri2}, we compute the average in \eqref{eqn:ave_1} and obtain that $
\dot {\tilde z}_i^a =\alpha \sum_{j \in \mathcal{N}_i}a_{ij}(\tilde z_j^a-\tilde z_i^a+v_j^a-v_i^a)-\kappa \beta \big(H_i (\tilde z_i^{a}+z^*)-b_i \big) $, which is uniform in $t_0$ and {independent of $e_i$.} Further, applying the similar technique to $\dot {v}_i $ leads to the following stochastic average system
\bee\label{eqn:error_ave}
\left\{
\begin{aligned}
\dot {\tilde z}_i^a&=\alpha \sum\nolimits_{j \in \mathcal{N}_i}a_{ij}(\tilde z_j^a-\tilde z_i^a+v_j^a-v_i^a)\\
&~~~-\kappa \beta \big(H_i (\tilde z_i^{a}+z^*)-b_i \big),\\
\dot {v}_i^a&=\sum\nolimits_{j \in \mathcal{N}_i}a_{ij}(\tilde z_i^a-\tilde z_j^a).\\
\end{aligned}
\right.
\ene
{\em Step 2: Stability of the stochastic average system \eqref{eqn:error_ave}.}

The stochastic average system of the networked $n$ vehicles can be compactly expressed as
\bee \label{eqn:ave_clo_n}
\left\{
\begin{aligned}
\dot{\tilde z}^{a}&=-\alpha (L \otimes I_m) (\tilde z^{a}+{v}^a)\\
&~~~~- \kappa\beta (H^d (\tilde z^{a}+1_n \otimes z^*)-b),\\ 
\dot {v}^a&=(L \otimes I_m) \tilde z^{a}.\\ 
\end{aligned}
\right.
\ene 
Denote an equilibrium point of \eqref{eqn:ave_clo_n} by $((\tilde z^{a}_{eq})',({v}^a_{eq})')'$. It follows that
\beq \label{eqn:v_eq}
\begin{array}{rl}
-\kappa\beta (H^d (\tilde z^{a}_{eq}+1_n \otimes z^*)-b) &= \alpha (L \otimes I_m) {v}^a_{eq},\\
(L \otimes I_m) \tilde z^{a}_{eq}&=0_{mn}.
\end{array}
\enq
Pre-multiplying both sides of the first equality of \eqref{eqn:v_eq} by $(1_n'\otimes I_m)$ and noting $1_n'L=0_{n}$, we obtain that  
\bee \label{eqilibia_1}
-\kappa\beta ({1_n'}\otimes I_m) (H^d (\tilde z^{a}_{eq}+1_n \otimes z^*)-b)= 0_{mn}.
\ene 
By the second equality of \eqref{eqn:v_eq} and Assumption \ref{assum_graph}, there exists a vector $z_1 \in\bR^m$ such that $\tilde z^{a}_{eq}=1_n \otimes z_1$. Inserting it into \eqref{eqilibia_1} yields that $ \big(\sum_{i=1}^nH_i\big)(z_1+z^*)=\sum_{i=1}^n b_i$. Since the source position $z^*$ in \eqref{eqn:opt1} satisfies that $z^*=\big(\sum_{i=1}^nH_i\big)^{-1}(\sum_{i=1}^n b_i)$ by Assumption \ref{assum_obj}, we obtain that $z_1=0_{m}$, i.e., $\tilde z^{a}_{eq}=0_{mn}$. That is, an equilibrium point of \eqref{eqn:ave_clo_n} is $(0'_{mn}, (v_{eq}^a)')'$.

Now, we {study the convergence of the networked average system \eqref{eqn:ave_clo_n}}. Defining $ V=\frac{1}{2}\big( \| \tilde z^{a}\|^2 + \alpha \cdot \| v^{a} - v_{eq}^a\|^2 \big) $ and taking its derivative along \eqref{eqn:ave_clo_n}, we obtain that
\bee \label{eqn:lya12}
\begin{aligned}
\dot V&= -\alpha ({\tilde z}^a)' (L \otimes I_m) (\tilde z^{a}+{v}^a) - \kappa\beta ({\tilde z}^a)'(H^d \tilde z^{a}\\
&~~~+H^d(1_n \otimes z^*) -b)+ \alpha(v^{a} - v_{eq}^a)'(L \otimes I_m) \tilde z^{a}.
\end{aligned}
\ene
By the first equality of \eqref{eqn:v_eq} and $\tilde z^{a}_{eq}=0$, $\dot V$ in \eqref{eqn:lya12} can be further simplified as
\bee \label{eqn:lya2}
\dot V=({\tilde z}^a)'(-\alpha (L \otimes I_m)-\kappa\beta H^d)\tilde z^{a}= -(\tilde z^{a})'M \tilde z^{a}.
\ene
{Clearly, the matrix $M$ is positive semi-definite, since $\alpha( L \otimes I_m) $ and $\kappa \beta H^d$ are positive semi-definite under Assumptions \ref{assum_graph} and \ref{assum_obj} \citep{ren2008distributed}}. Suppose that there exists a non-zero vector $\zeta \in \bR^{m n}$ such that $\zeta'M\zeta=0$. {That is, $ \zeta'\big(L \otimes I_m \big)\zeta=0$ and $\zeta'  H^d \zeta=0$. In light of Assumption \ref{assum_graph}, $\zeta$ must be of the form that $\zeta=1_n \otimes \zeta_1$ for some non-zero vector $\zeta_1 \in\bR^m$. Then, we can obtain that} ${\zeta_1'} \left(\sum_{i=1}^n H_i\right) \zeta_1=0$, which contradicts Assumption \ref{assum_obj} that $\sum_{i=1}^n H_i$ is strictly positive definite. Thus, the matrix $M$ is strictly positive definite.

{ Let $\lambda_1$ denote the smallest eigenvalue of $M$. It follows from \eqref{eqn:lya2} that $\dot V(t) \leq -\lambda_1 \| \tilde z^{a}(t) \|^2$.} Since $\frac{1}{2}\| \tilde z^{a}(t) \|^2 \leq V(t) = V(t_0)+\int_{t_0}^{t} \dot V(s) \text{d}s \leq V(t_0)+\int_{t_0}^{t}   -\lambda_1 \| \tilde z^{a}(s) \|^2 \text{d}s$, there exists a finite constant $\rho_1 >0$ such that
\bee \nonumber 
\|\tilde {z}_i^a(t) \| \leq \| \tilde z^{a} (t)\| \leq \rho_1\exp({-\lambda_1 t}).
\ene
Note that $\rho_1$ depends on the initial condition. 

{\em Step 3: Stability of {the dynamics} \eqref{eqn:error}.}

By \citet[Proposition 5.5]{oksendal2003stochastic} and Assumption \ref{assum_obj}, the dynamics \eqref{eqn:error} admits a unique (almost surely) continuous solution on $[t_0, +\infty)$. Together with \citet[Proposition 2]{liu2010stochastic}, the rest of proof is completed. 
\qed

{In addition, since $F(z_i)=F(z^*+\tilde z_i +\gamma \sin(\eta_i))$, it follows from \eqref{eqn:con_cor} that for any $\delta >0$,} $ \lim_{\epsilon \to 0} \cP\{|F(z_i)-F(z^*)| \leq O(\gamma)+O(\delta)+O( \exp({-\lambda_1 t})), \forall t \in [t_0,T_1(\epsilon)]\}=1.$

\subsection{ {Extension to directed graphs }}
For the case of directed graphs, the essential difference is that the Laplacian matrix $L$ is asymmetric and $\cG$ is unbalanced, i.e.,  $\sum_j a_{ij}\neq \sum_j a_{ji}$ for some $i\in\cV$. To solve it, the DSES controller in \eqref{eqn:control_1} is modified  as 
\bee \label{eqn:control_d}
\left \{
\begin{aligned}
u_i&=\alpha \sum\nolimits_{j \in \mathcal{N}_i}a_{ij}(\varphi(z_j-z_i)+v_j-v_i)
\\&~~~+{\beta}/{r_{ii}} \cdot \sin(\eta_i) \Delta_i +\gamma {\text{d} \text{sin}(\eta_i)}/{\text{d}t},\\
\dot v_i&= \sum\nolimits_{j \in \mathcal{N}_i}a_{ij}(z_i-z_j),\\
\dot {r}_i&= \sum\nolimits_{j \in \mathcal{N}_i}a_{ij}(r_j-r_i).
\end{aligned}
\right.
\ene
where $\alpha, \beta, \gamma, \varphi \in \bR_{+}$ are positive {control} parameters, $\varphi = \varrho+({1+\alpha})/({\alpha \varrho}) $ with {a constant} $\varrho >0$ and $r_i=(r_{1i}, r_{2i}, \ldots, r_{ni})' \in \bR^n$ with $r_{ii}(t_0)=1$, $r_{ji}(t_0)=0$ for $j \not = i$ is an estimate of a left eigenvector of $L$ associated with zero eigenvalue and can be computed in a distributed manner.

\begin{prop} \label{prop:integrator_d}
Given a directed graph $\cG$, consider the networked $n$ autonomous vehicles \eqref{eqn:integrator} under the DSES controller \eqref{eqn:control_d}. Let $\tilde{z}_i(t)=z_i(t) -\gamma \sin(\eta_i) -z^*$ where  $z^*$ is the source position defined in \eqref{eqn:opt1} and suppose that Assumptions \ref{assum_graph} and \ref{assum_obj} hold. Then, there exists a positive constant $\rho_2$ and a function $T_2(\epsilon): (0,\epsilon_0)\to \bN$ such that for any $\delta>0$ and bounded initial condition (i.e. $\twon{z_i(t_0)}< +\infty$, $\twon{v_i(t_0)}<+\infty$ for all $i\in\cV$), it holds that $\forall i \in \cV$,
\bee \nonumber
\lim_{\epsilon \to 0} \cP\{  \| \tilde z_i(t)\| \leq  \rho_2 \exp({- \lambda_2 t}) +\delta, \forall t \in [t_0,{T_2}(\epsilon)] \}=1,
\ene
where $\lim_{\epsilon\ra 0}T_2(\epsilon)=+\infty$ and $\lambda_2>0$ is given in \eqref{lambda2}. 
\end{prop}

{The proof and the explicit expression of $\lambda_2$ are given in Appendix \ref{proof_int_d}. Note that it usually holds that $\lambda_2< \lambda_1$,} i.e., the convergence rate here is slower than that in undirected graphs due to the directed interaction between vehicles.

\section{Illustrative Examples}
\label{sec:exam}

In this section, {examples are given to illustrate the effectiveness of the proposed DSES controllers.} Let $n=4$ in (\ref{eqn:opt1}) and
$$
\begin{aligned}
H_1&=\begin{bmatrix}   
    -2 & 1\\  
    1 & -0.5 
  \end{bmatrix},&
H_2=\begin{bmatrix}    
    -0.25 &0.5\\  
    0.5 & -1
  \end{bmatrix},\\
H_3&=\begin{bmatrix}   
    -0.5 & 1.5\\  
    1.5 & -4.5
  \end{bmatrix},&
H_4=\begin{bmatrix}    
    -3 & 1\\  
    1 & -0.33\end{bmatrix},
\end{aligned}
$$
$b_1=[1.5, -0.75]'$,  $b_2=[-0.5, 1]'$, $b_3=[-2, 6]'$, $b_4=[ 2.5, -0.83]'$, $c_1=0.44$, $c_2=0.5$, $c_3=-3$, $c_4=-0.042$. Note that $H_i$ is positive semi-definite, which clearly implies that $ \arg \max_{z} f_i(z) $ contains an infinite number of elements. Thus, a single vehicle is unable to guarantee to approach the source position $z^*=[1.53, 1.82]'$. 

\begin{figure}
  \centering
   \includegraphics[width=7cm]{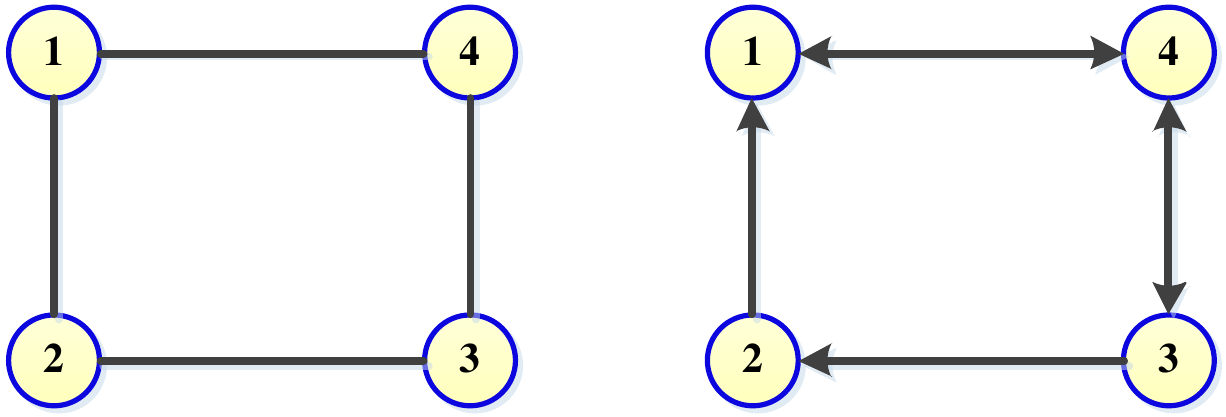}
   \caption{The undirected (left) and directed (right) interaction graphs of the multi-vehicle system.}
   \label{fig:graph}
\end{figure}

The interaction graphs among the vehicles are depicted in Fig. \ref{fig:graph}, which satisfy Assumption \ref{assum_graph}.
We select $\epsilon=0.05,$ $g=0.6,$ $ h=1$  for the excitation signals. Except Section \ref{sec:para}, the vehicles are initially placed at 
$z_1=[0, 0]', z_2=[0.9, 0]', z_3=[0.9, 0.9]',$ and $ z_4=[0, 0.9]'.$ 

\subsection{Simulations of undirected graphs}\label{subsec_simu}
\begin{figure}
    \centering
    \includegraphics[width=6cm]{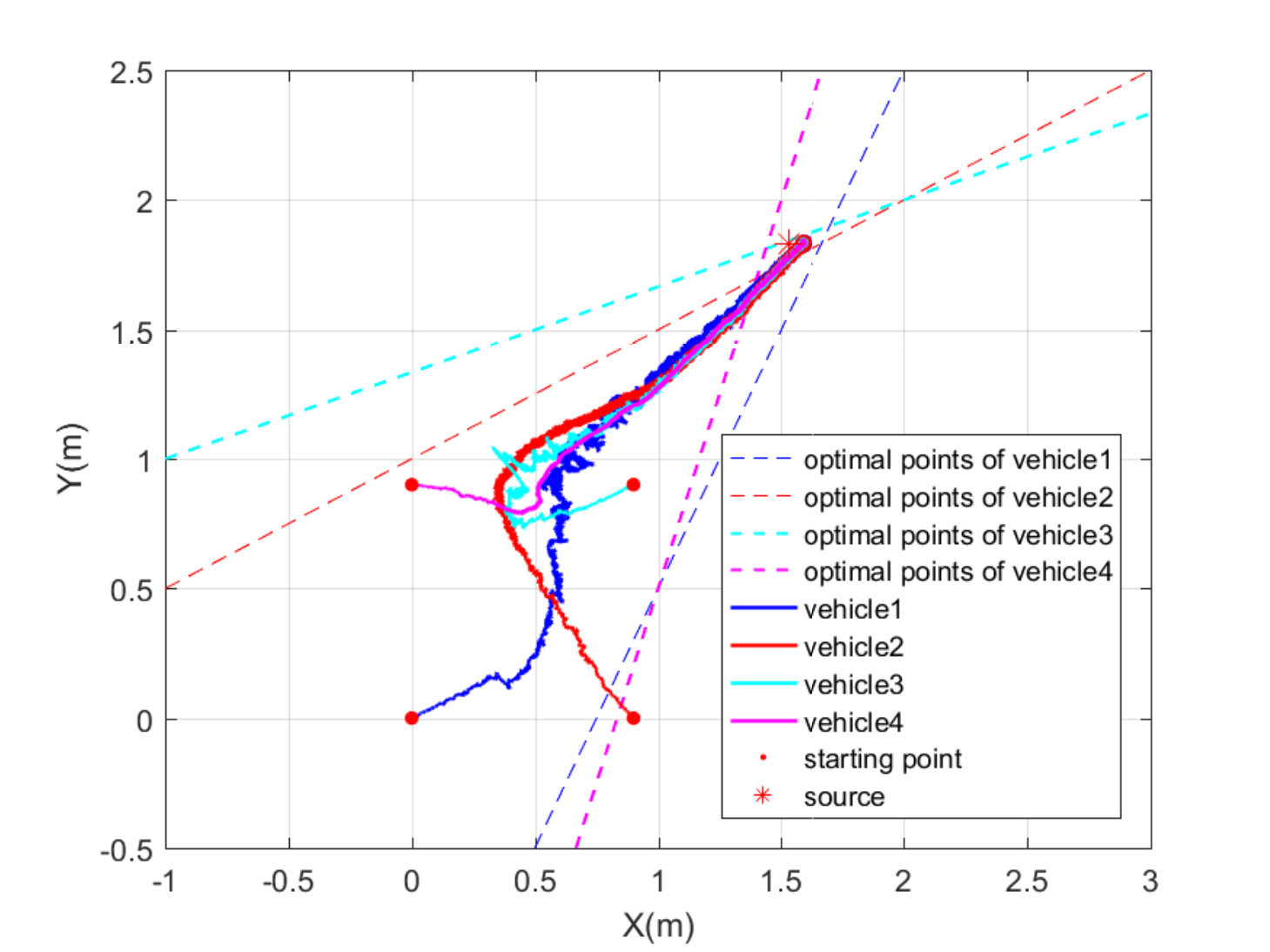}
    \caption{Trajectories of  vehicles under the DSES controller \eqref{eqn:control_1} in the undirected graph.}
    \label{fig:tra_int_v}
\end{figure}

\begin{figure}
  \centering
  \includegraphics[width=6cm]{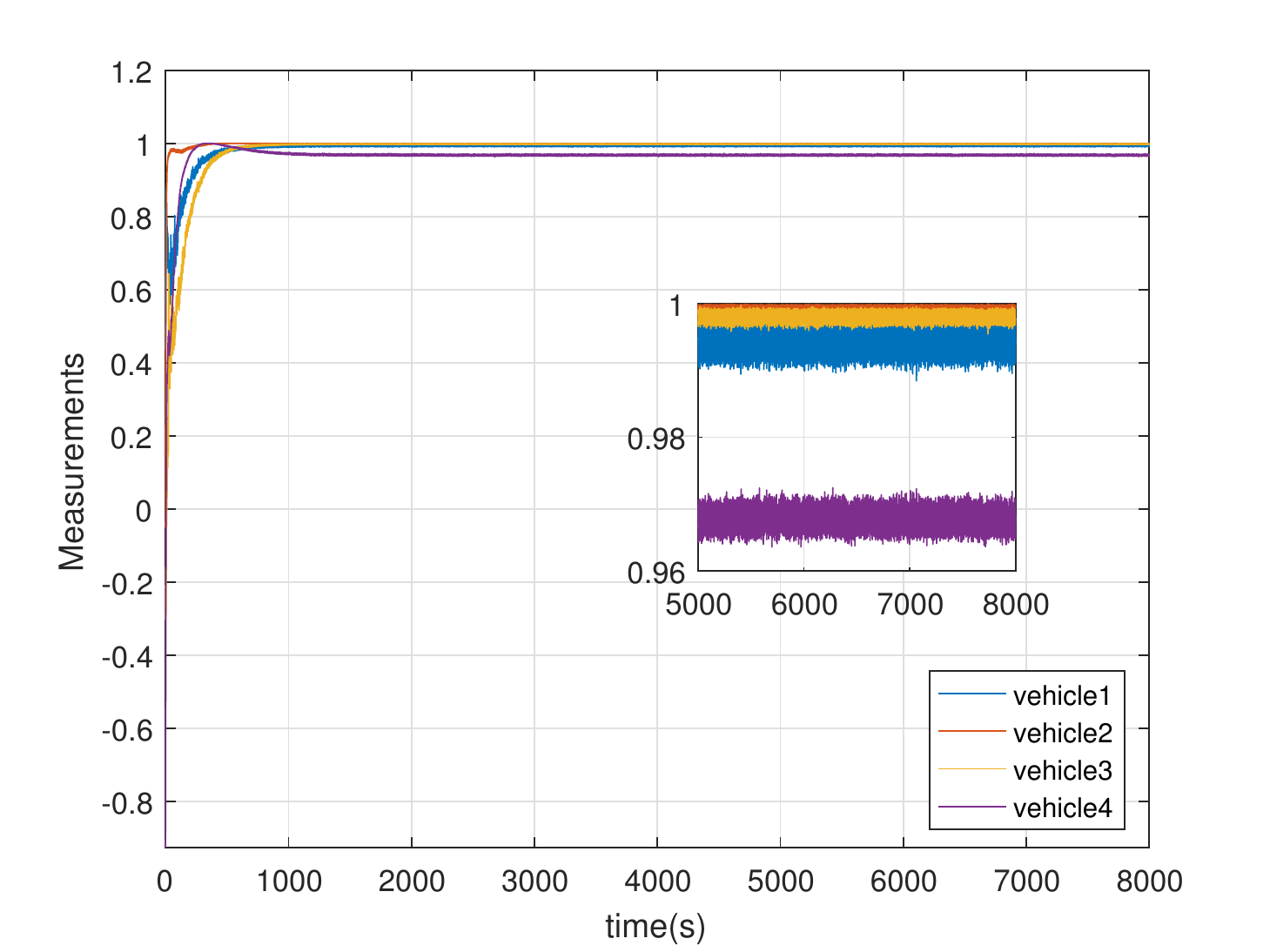}
  \caption{Measurement processes of  vehicles with the DSES controller \eqref{eqn:control_1} in the undirected graph.}
  \label{fig:J_int_v}
\end{figure}

Consider the multi-vehicle system \eqref{eqn:integrator} under the DSES controller \eqref{eqn:control_1} with $ \alpha=0.01,$ $ \beta=2.5$, and $ \gamma=0.01$. The results are shown in Fig. \ref{fig:tra_int_v}, where the solid lines denote the trajectories of the vehicles, and the dashed lines represent the optimal points of $f_i(z)$. Clearly, each vehicle is attracted to its own optimal points while tending to achieve consensus. Finally, all the vehicles revolve around the source position $z^*$, which is consistent with Proposition \ref{prop:int_v}, and the corresponding measurement processes are presented in Fig. \ref{fig:J_int_v}.

\begin{figure}
  \centering
    \includegraphics[width=6.5cm]{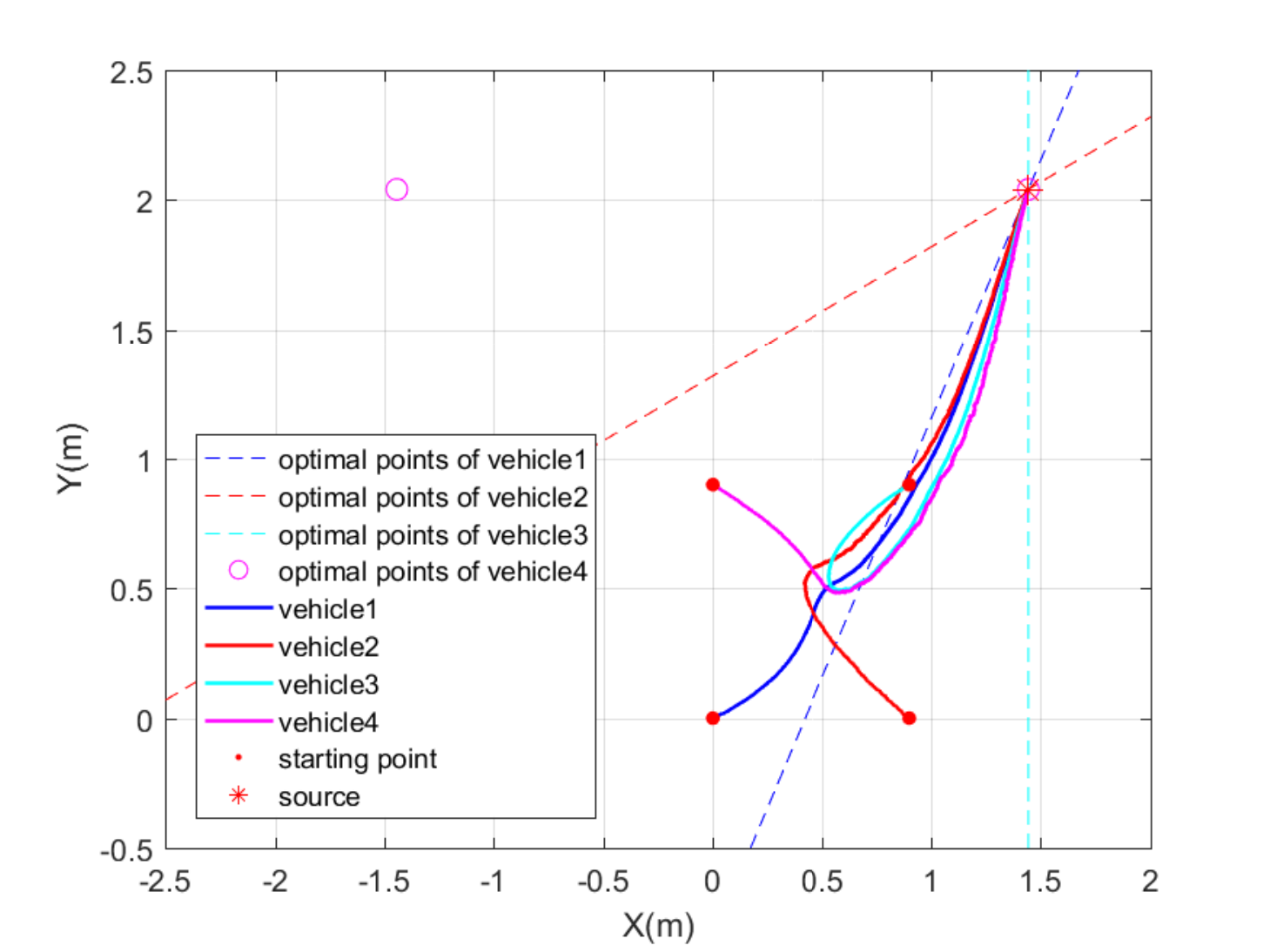}
    \caption{Trajectories of vehicles under the DSES controller \eqref{eqn:control_1} for non-quadratic $f_i(\cdot)$.}
    \label{fig:tra_nonqua}
\end{figure}

We also test our DSES controller (\ref{eqn:control_1}) for non-quadratic $f_i(\cdot)$, where $f_1, f_2$ remain quadratic with $H_1, H_2$ unchanged, $b_1=[2, -1]'$, $b_2=[-0.5, 1]'$, $c_1=-1$, $c_2=0$, and $f_3, f_4$ are redefined as non-quadratic
$$ 
\begin{aligned}
f_3&=0.083(x-2.44)^3-0.25(x-2.44)+0.83,\\
f_4&=-\text{e}^{-x^2- (y-1)^2}+2x^4\text{e}^{-x^2-(y-2)^2}-0.037.
\end{aligned}
$$
Thus, the optimal points of $f_3$ are two lines, i.e., $x=1.44$ and $x=3.44$ (we only draw the first one in Fig. \ref{fig:tra_nonqua}), $f_4$ have two isolated optimal points denoted by small magenta circles, and the optimal point of the aggregated objective function is $z^*=[1.443, 2.041]'$. Fig. \ref{fig:tra_nonqua} illustrates that the multi-vehicle system simultaneously approaches the source position, indicating that the cooperative source seeking method also works even for the non-quadratic case.

\subsection{Simulations of directed graphs} \label{subsec:simu_di}
\begin{figure}
  \centering
  \includegraphics[width=6cm]{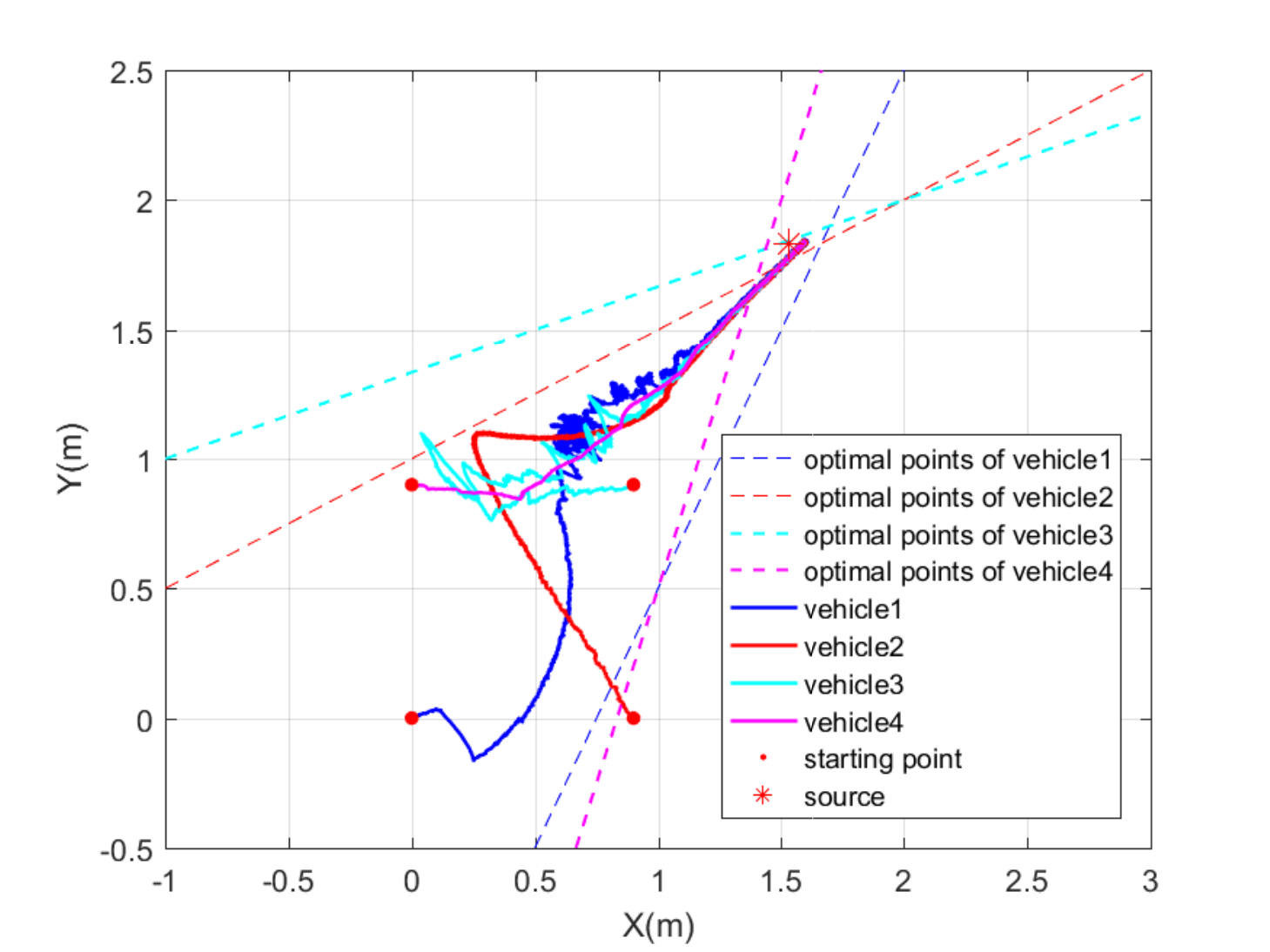}
  \caption{Trajectories of vehicles under the DSES controller \eqref{eqn:control_d} in the directed graph.}
  \label{fig:tra_int_d}
\end{figure}

\begin{figure}
  \centering
  \includegraphics[width=6cm]{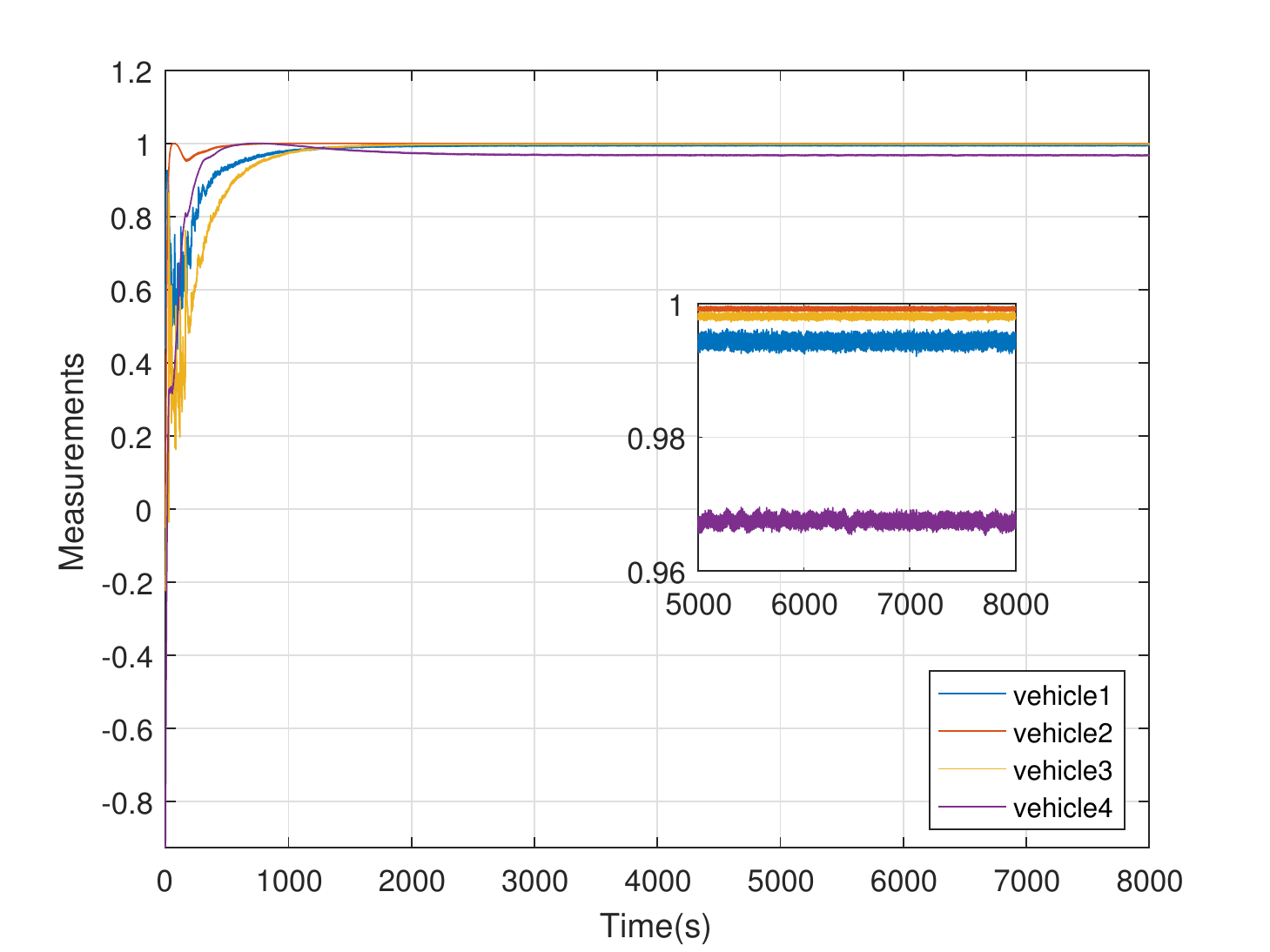}
  \caption{Measurement processes of vehicles with the DSES controller \eqref{eqn:control_d} in the directed graph.}
  \label{fig:J_int_d}
\end{figure}

Consider the vehicles in directed interaction graphs. The control parameters of the DSES controller \eqref{eqn:control_d} are set as $ \alpha=0.002,$ $ \beta=1.2$, $ \gamma=0.0125$ and { $\varphi=50$.}
Fig. \ref{fig:tra_int_d} shows that all vehicles converge to the neighborhood of the source position $z^*$, and Fig. \ref{fig:J_int_d} shows the measurement processes. {Comparing Fig. \ref{fig:J_int_d} with Fig. \ref{fig:J_int_v}, we can see the convergence rate of the directed graph is slower than that of the undirected graph. This is reasonable since there are fewer interaction links in the directed graph than in the undirected graph.}

\subsection{Effects of control parameters}
\label{sec:para}

In this subsection, we illustrate the effects of the two parameters $\alpha$ and $\beta$ by the DSES controller (\ref{eqn:control_1}) for the networked vehicles in the undirected graph. First, let the initial positions $z_1= z_2 =z_3= z_4=[0.5, 0.5]'$ and $\alpha=0.005$. This indicates that all vehicles start {from a consensus position}, and $\alpha$ in Section \ref{subsec_simu} is reduced by one half.  Then, let the initial positions $z_1=[0.75, 0]', z_2=[0.5, 1.25]', z_3=[0.75, 1.58]', z_4=[1, 0.5]'$ and $ \beta=1.25$. One can verify that each vehicle starts from an optimal point of its local objective function, and $\beta$ in Section \ref{subsec_simu} is reduced by one half. 

\begin{figure}
  \centering
  \includegraphics[width=6cm]{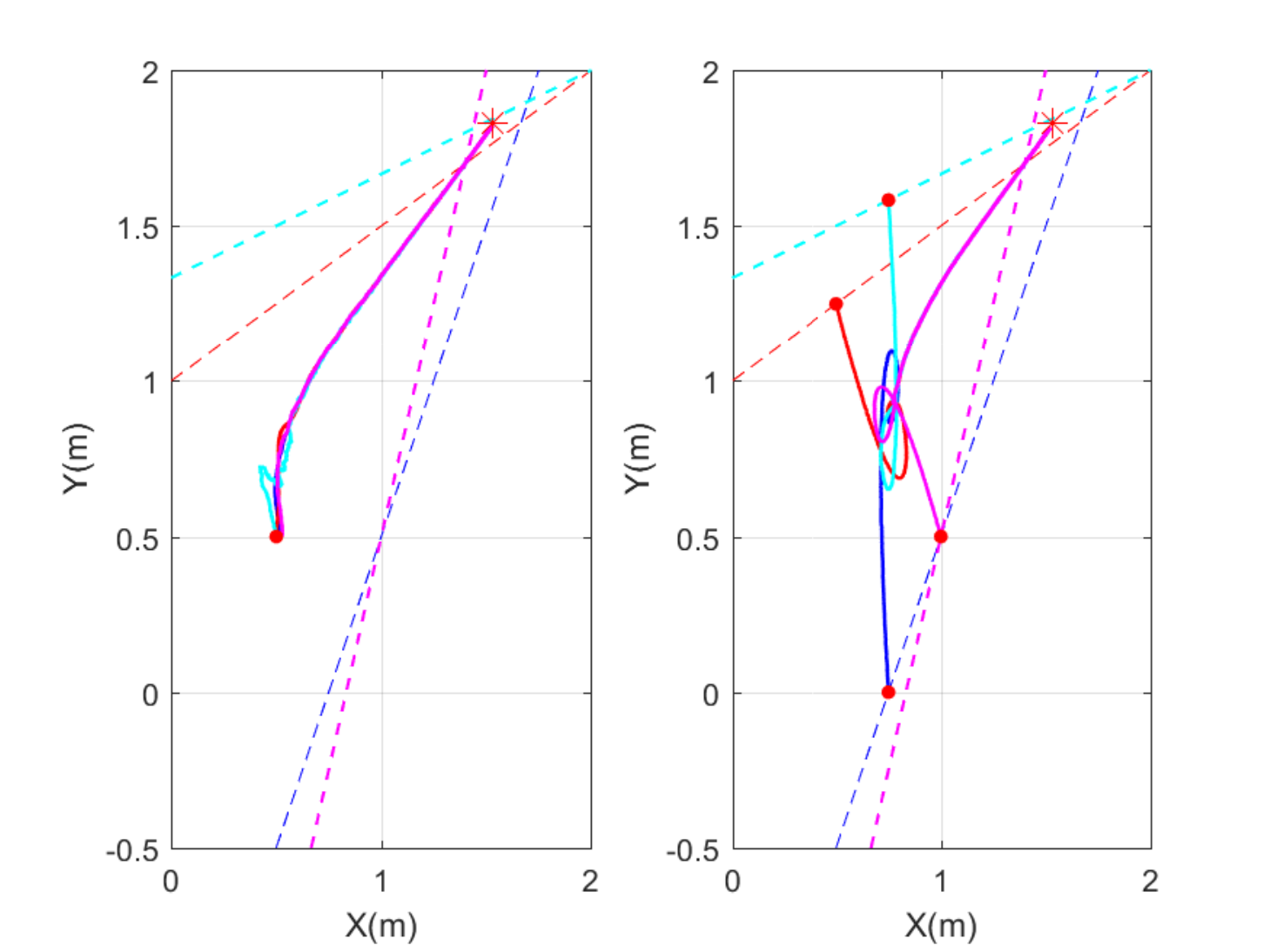}
  \caption{Trajectories of vehicles starting from {a consensus position} (left) and optimal points of $f_i(z)$ (right).}
  \label{fig:tra_con_com}
\end{figure}

\begin{figure}
  \centering
  \includegraphics[width=6cm]{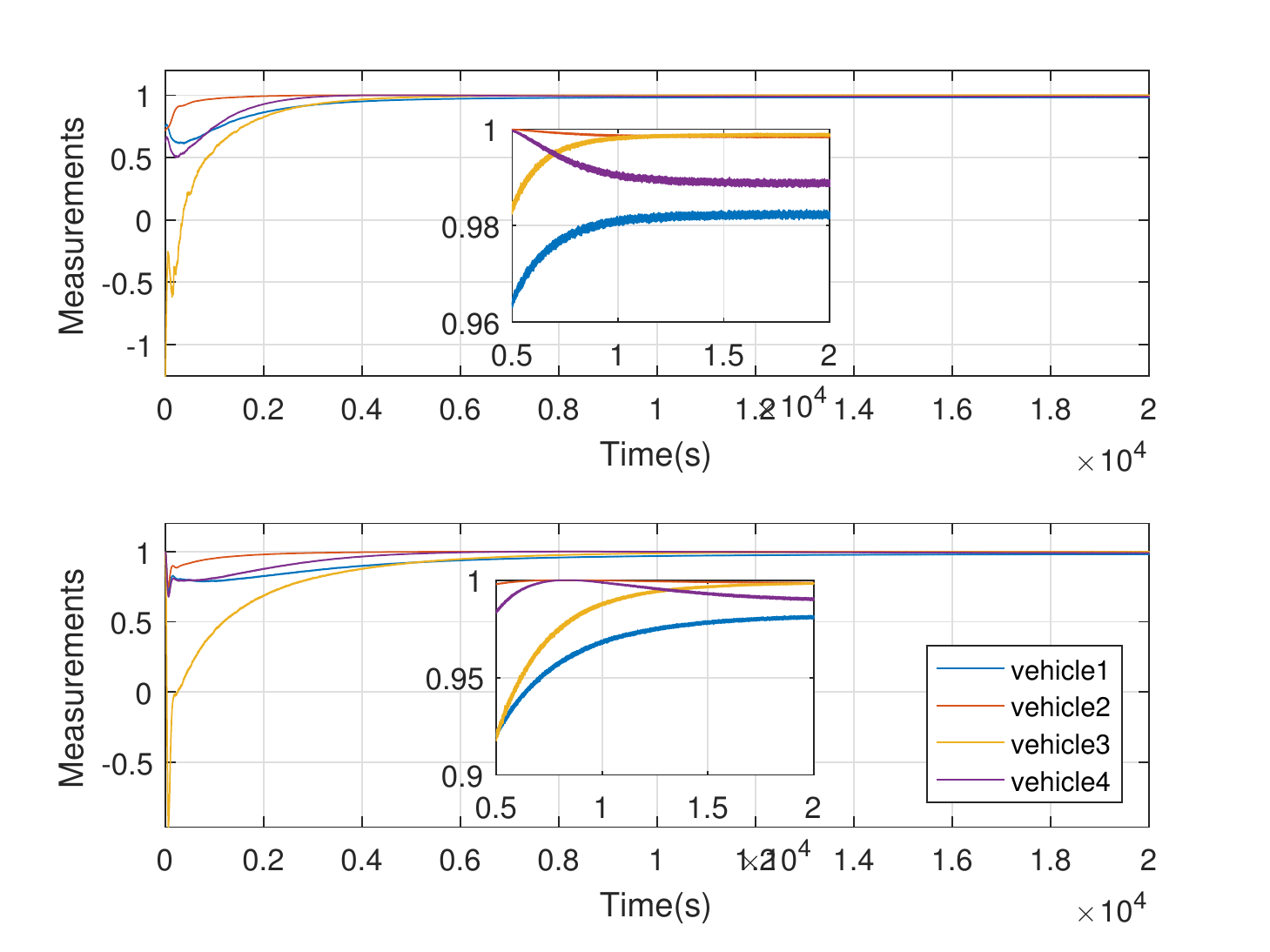}
  \caption{Measurement processes of vehicles starting from {a consensus position} (top) and optimal points of $f_i(z)$ (bottom).}
  \label{fig:J_con_com}
\end{figure}

The comparisons of the two cases are shown in Fig. \ref{fig:tra_con_com} and \ref{fig:J_con_com}. {In the left subfigure of Fig. \ref{fig:tra_con_com}, each vehicle tends to approach a consensus state, while it moves to its local optimal point in the right subfigure.} That is, the consensus term forces the vehicles to reach consensus and the stochastic ES term for gradient estimation drives the vehicles to their own optimal points. Both objectives are essential to approach the source position. 

\section{Conclusion}
\label{sec:conclusion}
We have proposed the {DSES} controllers of the multi-vehicle system for the cooperative source seeking problem. In our scheme, each vehicle is only required to obtain the values of its local objective function and the relative positions to its neighbors. Via the stochastic averaging theory, we establish the convergence of the networked vehicles under our DSES controllers in probability, in both undirected and directed interaction graphs. Finally, simulations are included to verify our theoretical results.

\appendix
\section{Proof of Proposition \ref{prop:integrator_d}}\label{proof_int_d}

In the appendix, the concatenated vector of two vectors $x$ and $y$ is simply written as $[x;y]=(x',y')'$.

Consider the autonomous vehicle \eqref{eqn:integrator} under the DSES controller \eqref{eqn:control_d}, it holds that
\beq \label{eqn:err_d}
\left\{\begin{aligned}
\dot {\tilde{z}}_i&=\alpha \sum\nolimits_{j \in \mathcal{N}_i}a_{ij}(\varphi(\tilde z_j+\gamma \text{sin}(\eta_j)- \tilde z_i-\gamma \text{sin}(\eta_i))
\\&~~~~~~~~~~~~~~~~~~+v_j-v_i)+{\beta}/{r_{ii}} \cdot \sin(\eta_i) \Delta_i, \\
\dot v_i&= \sum\nolimits_{j \in \mathcal{N}_i}a_{ij}(\tilde z_i+\gamma \text{sin}(\eta_i)-\tilde z_j-\gamma \text{sin}(\eta_j)),\\
\dot r_i&= \sum\nolimits_{j \in \mathcal{N}_i}a_{ij}(r_j-r_i).
\end{aligned} \right.
\enq
Similar to the proof of Proposition \ref{prop:int_v}, we define the stochastic average system of the networked multi-vehicle system, apply the same technique as in \eqref{eqn:ergodic}-\eqref{eqn:tri2}, and obtain that
\bee \label{eqn:ave_dg}
\left\{
\begin{aligned}
\dot{\tilde z}^{a}&=-\alpha (L \otimes I_m) ( \varphi \tilde z^{a} + {v}^a) - \kappa\beta(R^{-1} \otimes I_m)\\
&~~~~ (H^d  (\tilde z^{a}+1_n \otimes z^*) -b),\\ 
\dot {v}^a&=(L \otimes I_m) \tilde z^{a},\\ 
\dot {r}&=-(L \otimes I_n) r,
\end{aligned}
\right.
\ene 
where  $R=\diag(r_{11}, \ldots, r_{nn})$. 

To study the {convergence} of \eqref{eqn:ave_dg}, we first consider its approximated system below
\bee \label{eqn:ave_dg_sim1}
\left\{
\begin{aligned}
\dot{\tilde z}^{a}&=-\alpha (L \otimes I_m) ( \varphi \tilde z^{a} + {v}^a)- \kappa\beta( \Xi ^{-1} \otimes I_m) \\
&~~~~(H^d  (\tilde z^{a}+1_n \otimes z^*)-b),\\ 
\dot {v}^a&=(L \otimes I_m) \tilde z^{a},
\end{aligned}
\right.
\ene 
where $R$ in \eqref{eqn:ave_dg} is replaced by $\Xi=\diag(\xi)$ with $\xi \in \bR^{1 \times n}_+$ being a positive left eigenvector of $L$ associated with zero eigenvalue \citep[Lemma 1]{zhu2018continuous}.
Let $[\tilde z_{eq}^a; v_{eq}^a]$ be an equilibrium point of \eqref{eqn:ave_dg_sim1}. Then we have that
 \bee \label{eq}
 -\kappa\beta ( \Xi^{-1} \otimes I_m) (H^d  (\tilde z^a_{eq} +1_n \otimes z^*)-b)= \alpha  (L \otimes I_m) v_{eq}^a. 
\ene
Since ${1_n'} \Xi =\xi $, we pre-multiply both sides by ${1_n'} \Xi \otimes I_m$ and {obtain the same relation as that in \eqref{eqilibia_1}.}
Then, we can obtain $\tilde z^a_{eq}=0_{mn}$ as in the proof of Proposition \ref{prop:int_v}. 

Consider the following Lyapunov functional candidate for the system \eqref{eqn:ave_dg_sim1} 
\bee \label{eqn:ly_dg}
\begin{aligned}
V(x)&=\frac{1}{2}x'\left(\begin{bmatrix}
  			 \varrho^2+1 & \varrho \\
  			\varrho & 1
 			 \end{bmatrix}   \otimes (\Xi \otimes I_m)\right)x,
\end{aligned}
\ene
where $x=[\tilde z^a; v^a- v_{eq}^a ]$. Clearly, it holds that
\bee \label{eqn:vd_con1}
a_1 \|x\|^2 \leq V(x) \leq a_2\|x\|^2,  \|\nabla V(x)\| \leq 2a_2 \|x\|,
\ene 
where $\nabla V(x)$ is the gradient of $V(x)$, and the positive constant $a_i, i\in\{1,2\}$ is not explicitly given for brevity. 
{We take derivative of \eqref{eqn:ly_dg} along \eqref{eqn:ave_dg_sim1} and obtain that} $ \dot V=-x'(P \otimes(\Xi L+L' \Xi)\otimes I_m+Q\otimes H^d)x$ with
$$P= \frac{\alpha}{2} \begin{bmatrix}  \varrho^3+2 \varrho +\frac{\alpha+1}{ \alpha \varrho} & 1+\varrho^2 \\ 1+\varrho^2& \varrho \end{bmatrix} \text{and}~ Q=\kappa \beta\begin{bmatrix} 1+\varrho^2 & \frac{\varrho}{2}\\ \frac{\varrho}{2} & 0\end{bmatrix}$$
{by the selection of $\varphi = \varrho+ ({1+\alpha})/({\alpha \varrho}) $ and the relation in \eqref{eq}.} 
Since $\xi L=0'_{n}$, {pre-multiplying both sides of the second equality in \eqref{eqn:ave_dg_sim1} by $\xi \otimes I_m$ yields that} $(\xi \otimes I_m)\dot {v}^a=0_m$, which implies that  $(\xi \otimes I_m)(v^a- v_{eq}^a)=0_m$.  Hence we only need to investigate the convergence of \eqref{eqn:ave_dg_sim1} on the subspace 
$\cX=\{ [x_1;x_2] \in \bR^{2mn}| (\xi \otimes I_m) x_2=0_m \}.$ 

Let $\cX_1=\{ [x_1;x_2] \in \bR^{2mn}| x_1=1_n \otimes \zeta, x_2= 0_{mn}, \zeta \in \bR^m \} $ 
{denote the null space of $P \otimes(\Xi L+L' \Xi)\otimes I_m$ on $\cX$}. Then, for any $x = [1_n \otimes \zeta; 0_{mn}] \in\cX_1$, it holds that
\begin{align} \label{space1}
\dot V & =-[1_n \otimes \zeta;0]'(Q \otimes H^d ) [1_n \otimes \zeta;0] \nonumber \\
&= -\kappa \beta (1+\varrho^2) \zeta ' \big( \sum_{i=1}^n H_i \big) \zeta\leq -c_0 \twon{x}^2
\end{align}
where $c_0$ is positive by Assumption \ref{assum_obj}.  
For any $x  \in \cX-\cX_1$, applying Weyl's inequality \citep[Theorem 4.3.1] {horn2012matrix} yields that 
\begin{align} 
\dot V &=-x'(P \otimes(\Xi L+L' \Xi)\otimes I_m+ Q \otimes H^d)x  \nonumber\\
& \leq - (\lambda_P \lambda_L-\lambda_Q \lambda_{H}) \| x \|^2, \label{space2}
\end{align}
where $\lambda_P=O(\varrho)$ is the smallest eigenvalue of $P$, $\lambda_L$ is the smallest nonzero eigenvalue of $\Xi L+L' \Xi$, $\lambda_Q =O(\varrho^2)$ is the largest eigenvalue of $-Q$, and $\lambda_{H}$ is the largest eigenvalue of $H^d$. Since $\lim_{\varrho\ra 0}\lambda_P/\lambda_Q = +\infty$, there exists a positive $\varrho_0$ such that $\lambda_P \lambda_L-\lambda_Q \lambda_{H}>0$ for any $\varrho \in (0, \varrho_0)$. Jointly with \eqref{eqn:vd_con1}, \eqref{space1} and \eqref{space2}, it holds that  $\forall x \in \cX$ and $\varrho \in (0, \varrho_0)$, there exists a positive $a_3>0$ such that
\bee \label{eqn:vd_con2}
\dot V \leq -a_3 V,
\ene 
Therefore, $[0_{mn}; v_{eq}^a]$ is an exponentially stable equilibrium point of \eqref{eqn:ave_dg_sim1}. 

Now, we consider the dynamics of ${\tilde z}^{a}$ in \eqref{eqn:ave_dg}  and rewrite it as
\bee \label{pertubated}
\begin{aligned}
\dot{\tilde z}^{a}&=-\alpha (L \otimes I_m) ( \varphi \tilde z^{a} + {v}^a) - \kappa\beta( \Xi^{-1} \otimes I_m)\\
&~~~~ (H^d ( \tilde z^{a} + 1_n \otimes z^*) -b)+d,\\ 
\end{aligned}
\ene 
where $d=-\kappa\beta (R^{-1}-\Xi^{-1} ) \otimes I_m (H^d ( \tilde z^{a} + 1_n \otimes z^*)-b)$ is treated as a perturbation term of \eqref{eqn:ave_dg_sim1} and satisfies that
\beq \label{eqn:disturb}
\begin{aligned}
\| d \|&\leq \| \kappa \beta (R^{-1}-\Xi^{-1}) \otimes I_m H^d \tilde z^{a} \| \\
&~~~+ \| \kappa\beta (R^{-1}-\Xi^{-1} ) \otimes I_m (H^d (1_n \otimes z^*)-b) \| \\
&\leq \kappa \beta \varsigma(t)\|H^d \tilde z^{a} \|+ \kappa\beta \varsigma(t) \| (H^d (1_n \otimes z^*)-b) \|,
\end{aligned}
\enq
where $\varsigma(t) = \| R^{-1}-\Xi^{-1}\| \otimes I_m $. In view of Lemma 1 in \citet{zhu2018continuous},  $\|r(t)-1_n \otimes \xi' \|$ converges to zero less slowly than the rate of $\exp(-\ell t)$, where $\ell$ is the real part of the eigenvalue of $L$ that is closest to the left half plane and is positive under Assumption \ref{assum_graph}. This implies that $\varsigma(t)$ tends to zero at the same rate,
\bee\label{pertu}
\sup_{t \ge t_0} \varsigma(t) < +\infty ~\text{and}~ \int_{t_0}^ {+\infty}  \varsigma(s) \text{d}s  < a_4
\ene
for some $a_4 >0$. Note that \eqref{eqn:ave_dg_sim1} has an exponentially stable equilibrium point $[0; v_{eq}^a]$, and its Lyapunov function \eqref{eqn:ly_dg} satisfies \eqref{eqn:vd_con1} and \eqref{eqn:vd_con2} in $\cX$. {Provided the perturbation term $d(t)$ satisfies \eqref{eqn:disturb} and \eqref{pertu}, for any bounded initial condition of \eqref{pertubated},} there exists a positive $\rho_2$ such that $\|\tilde z^a\|  \leq \frac{\rho_2}{2} \big(\exp(- a_3 t)+\exp{(-\ell t})\big) $ \citep[Lemma 9.4]{khalil2002nonlinear}, which further implies that
\bee \label{lambda2}
\|\tilde z_i^a\| \leq \rho_2 \exp(-\min \{{a_3}, \ell\}t)  :=\rho_2 \exp (-\lambda_2 t).
\ene
The rest of the proof is similar to Step 3 in Proposition \ref{prop:int_v} and is omitted.\qed

\bibliographystyle{agsm} 
\bibliography{RefDBase} 

@article{guay2018distributed,
	Author = {Guay, Martin and Vandermeulen, Isaac and Dougherty, Sean and McLellan, P James},
	Journal = {Automatica},
	Pages = {498--509},
	Publisher = {Elsevier},
	Title = {Distributed extremum-seeking control over networks of dynamically coupled unstable dynamic agents},
	Volume = {93},
	Year = {2018}}

@article{you2019distributed,
	Author = {You, Keyou and Tempo, Roberto and Xie, Pei},
	Journal = {IEEE Transactions on Automatic Control},
	Number = {3},
	Pages = {880--895},
	Publisher = {IEEE},
	Title = {Distributed algorithms for robust convex optimization via the scenario approach},
	Volume = {64},
	Year = {2019}}

@inproceedings{dhariwal2004bacterium,
	Author = {Dhariwal, Amit and Sukhatme, Gaurav S and Requicha, Aristides AG},
	Booktitle = {International Conference on Robotics and Automation (ICRA)},
	Organization = {IEEE},
	Pages = {1436--1443},
	Title = {Bacterium-inspired robots for environmental monitoring},
	Year = {2004}}

@article{gao2016detection,
	Author = {Gao, Xiang and Acar, Levent and Sarangapani, Jaganathan},
	Journal = {Journal of Automation and Control Engineering},
	Number = {6},
	Pages = {418--423},
	Title = {Detection and Tracking of Odor source in Sensor Networks Using Reasoning System},
	Volume = {4},
	Year = {2016}}

@phdthesis{zhao2016acoustic,
	Author = {Zhao, Nilu},
	School = {Massachusetts Institute of Technology},
	Title = {Acoustic source localization},
	Year = {2016}}

@article{gao20163d,
	Author = {Gao, Bo and Li, Hongbo and Li, Wei and Sun, Fuchun},
	Journal = {Adaptive Behavior},
	Number = {1},
	Pages = {52--65},
	Publisher = {SAGE Publications Sage UK: London, England},
	Title = {3{D} Moth-inspired chemical plume tracking and adaptive step control strategy},
	Volume = {24},
	Year = {2016}}

@inproceedings{turgeman2018multiple,
	Author = {Turgeman, Avi and Werner, Herbert},
	Booktitle = {American Control Conference (ACC)},
	Organization = {IEEE},
	Pages = {3558--3563},
	Title = {Multiple Source Seeking using Glowworm Swarm Optimization and Distributed Gradient Estimation},
	Year = {2018}}

@article{bri2016distributed,
	Author = {Brinon-Arranz, Lara and Schenato, Luca and Seuret, Alexandre},
	Journal = {IEEE Transactions on Control of Network Systems},
	Number = {2},
	Pages = {104--115},
	Title = {Distributed Source Seeking via a Circular Formation of Agents Under Communication Constraints},
	Volume = {3},
	Year = {2016}}

@article{khong2014multi,
	Author = {Khong, Sei Zhen and Tan, Ying and Manzie, Chris},
	Journal = {Automatica},
	Number = {9},
	Pages = {2312--2320},
	Title = {Multi-agent source seeking via discrete-time extremum seeking control},
	Volume = {50},
	Year = {2014}}

@article{durr2017extremum,
	Author = {D{\"u}rr, Hans Bernd and Krsti{\'c}, Miroslav and Scheinker, Alexander and Ebenbauer, Christian},
	Journal = {Automatica},
	Pages = {91--99},
	Title = {Extremum seeking for dynamic maps using Lie brackets and singular perturbations},
	Volume = {83},
	Year = {2017}}

@article{manzie2009extremum,
	Author = {Manzie, Chris and Krstic, Miroslav},
	Journal = {IEEE Transactions on Automatic Control},
	Number = {3},
	Pages = {580--585},
	Publisher = {IEEE},
	Title = {Extremum seeking with stochastic perturbations},
	Volume = {54},
	Year = {2009}}

@article{liu2010stochastic,
	Author = {Liu, Shujun and Krstic, Miroslav},
	Journal = {IEEE Transactions on Automatic Control},
	Number = {10},
	Pages = {2235--2250},
	Publisher = {IEEE},
	Title = {Stochastic averaging in continuous time and its applications to extremum seeking},
	Volume = {55},
	Year = {2010}}

@article{frihauf2014single,
	Author = {Frihauf, Paul and Liu, Shujun and Krstic, Miroslav},
	Journal = {Journal of Dynamic Systems, Measurement, and Control},
	Number = {5},
	Pages = {051024},
	Publisher = {American Society of Mechanical Engineers},
	Title = {A single forward-velocity control signal for stochastic source seeking with multiple nonholonomic vehicles},
	Volume = {136},
	Year = {2014}}

@article{zhang2016extremum,
  title={Extremum seeking control of a nonholonomic system with sensor constraints},
  author={Zhang, Yinghua and Makarenkov, Oleg and Gans, Nicholas},
  journal={Automatica},
  volume={70},
  pages={86--93},
  year={2016},
  publisher={Elsevier}
}

@article{vandermeulen2018discrete,
	Author = {Vandermeulen, Isaac and Guay, Martin and McLellan, P James},
	Journal = {IEEE Transactions on Control of Network Systems},
	Number = {3},
	Pages = {1182--1192},
	Publisher = {IEEE},
	Title = {Discrete-time distributed extremum-seeking control over networks with unstable dynamics},
	Volume = {5},
	Year = {2018}}

@article{dougherty2017extremum,
	Author = {Dougherty, Sean and Guay, Martin},
	Journal = {IEEE Transactions on Automatic Control},
	Number = {2},
	Pages = {928--933},
	Publisher = {IEEE},
	Title = {An extremum-seeking controller for distributed optimization over sensor networks},
	Volume = {62},
	Year = {2017}}

@article{ye2016distributed,
	Author = {Ye, Maojiao and Hu, Guoqiang},
	Journal = {IEEE Transactions on Control Systems Technology},
	Number = {6},
	Pages = {2048--2058},
	Publisher = {IEEE},
	Title = {Distributed extremum seeking for constrained networked optimization and its application to energy consumption control in smart grid},
	Volume = {24},
	Year = {2016}}

@article{li2015distributed,
  title={Distributed extremum seeking and formation control for nonholonomic mobile network},
  author={Li, Chaoyong and Qu, Zhihua and Weitnauer, Mary Ann},
  journal={Systems \& Control Letters},
  volume={75},
  pages={27--34},
  year={2015},
  publisher={Elsevier}
}

@inproceedings{li2018distributed,
	Author = {Li, Zhuo and You, Keyou and Song, Shiji and Dong, Siqi},
	Booktitle = {International Conference on Control and Automation (ICCA)},
	Organization = {IEEE},
	Pages = {367--372},
	Title = {Distributed Extremum Seeking with Stochastic Perturbations},
	Year = {2018}}

@article{lin2017stochastic,
	Author = {Lin, Jinbiao and Song, Shiji and You, Keyou and Krstic, Miroslav},
	Journal = {Automatica},
	Pages = {378--386},
	Publisher = {Elsevier},
	Title = {Stochastic source seeking with forward and angular velocity regulation},
	Volume = {83},
	Year = {2017}}

@inproceedings{poveda2013distributed,
	Author = {Poveda, J and Quijano, N},
	Booktitle = {American Control Conference (ACC)},
	Organization = {IEEE},
	Pages = {2772-2777},
	Title = {Distributed extremum seeking for real-time resource allocation},
	Year = {2013}}

@inproceedings{menon2014collaborative,
	Author = {Menon, Anup and Baras, John S},
	Booktitle = {Conference on Decision and Control (CDC)},
	Organization = {IEEE},
	Pages = {346--351},
	Title = {Collaborative extremum seeking for welfare optimization},
	Year = {2014}}

@book{ren2008distributed,
	Author = {Ren, Wei and Beard, Randal W},
	Publisher = {Springer},
	Title = {Distributed consensus in multi-vehicle cooperative control},
	Year = {2008}}

@book{oksendal2003stochastic,
	Author = {{\O}ksendal, Bernt},
	Publisher = {Springer},
	Title = {Stochastic differential equations},
	Year = {2003}}

@book{chen1998linear,
	Author = {Chen, Chi Tsong},
	Publisher = {Oxford University Press, Inc.},
	Title = {Linear system theory and design},
	Year = {1998}}

@article{gharesifard2014distributed,
	Author = {Gharesifard, Bahman and Cort{\'e}s, Jorge},
	Journal = {IEEE Transactions on Automatic Control},
	Number = {3},
	Pages = {781--786},
	Publisher = {IEEE},
	Title = {Distributed continuous-time convex optimization on weight-balanced digraphs},
	Volume = {59},
	Year = {2014}}

@inproceedings{wang2010control,
	Author = {Wang, Jing and Elia, Nicola},
	Booktitle = {Allerton Conference on Communication, Control, and Computing (Allerton)},
	Organization = {IEEE},
	Pages = {557--561},
	Title = {Control approach to distributed optimization},
	Year = {2010}}

@book{ash2000probability,
	Author = {Ash, Robert B and Dol{\'e}ans-Dade, Catherine A},
	Publisher = {Academic Press},
	Title = {Probability and measure theory},
	Year = {2000}}

@article{zhu2018continuous,
	Author = {Zhu, Yanan and Yu, Wenwu and Wen, Guanghui and Ren, Wei},
	Journal = {IEEE Transactions on Circuits and Systems II: Express Briefs},
	Number = {7},
	Pages = {1202--1206},
	Publisher = {IEEE},
	Title = {Continuous-time coordination algorithm for distributed convex optimization over weight-unbalanced directed networks},
	Volume = {66},
	Year = {2019}}

@book{khalil2002nonlinear,
	Author = {Khalil, Hassan K},
	Publisher = {Prentice-Hall},
	Title = {Nonlinear systems},
	Year = {2002}}

@book{horn2012matrix,
	Author = {Horn, Roger A and Johnson, Charles R},
	Publisher = {Cambridge university press},
	Title = {Matrix analysis},
	Year = {2012}}

@inproceedings{kvaternik2012analytic,
  title={An analytic framework for decentralized extremum seeking control},
  author={Kvaternik, Karla and Pavel, Lacra},
  booktitle={American Control Conference (ACC)},
  pages={3371--3376},
  year={2012},
  organization={IEEE}
}

@inproceedings{michalowsky2017distributed,
  title={Distributed extremum seeking over directed graphs},
  author={Michalowsky, Simon and Gharesifard, Bahman and Ebenbauer, Christian},
  booktitle={Conference on Decision and Control (CDC)},
  pages={2095--2101},
  year={2017},
  organization={IEEE}
}

\end{document}